\documentclass[aps,prl,twocolumn,superscriptaddress]{revtex4-1}
\usepackage{amsmath,verbatim}
\usepackage[capitalise]{cleveref}
\usepackage{tabularx, threeparttable,multirow,graphicx,xcolor,soul,float}

\begin{document}


\title{Beta-decay Half Lives beyond $^{54}$Ca: A Systematic Survey of Decay Properties approaching the Neutron Dripline}


\author{W.-J. Ong}
\affiliation{Lawrence Livermore National Laboratory, Livermore CA 94550, USA}
\author{Z. Y. Xu}
\affiliation{Department of Physics and Astronomy, University of Tennessee, Knoxville, TN 37996, USA}
\author{R. Grzywacz}
\affiliation{Department of Physics and Astronomy, University of Tennessee, Knoxville, TN 37996, USA}
\author{A. Ravli\'{c}}
\affiliation{Facility for Rare Isotope Beams, Michigan State University, East Lansing, MI 48824, USA}
\affiliation{Department of Physics, Faculty of Science, University of Zagreb, Bijeni\v cka c. 32, 10000 Zagreb, Croatia}
\author{I. Cox}
\affiliation{Department of Physics and Astronomy, University of Tennessee, Knoxville, TN 37996, USA}
\author{J.~M. Allmond}
\affiliation{Physics Division, Oak Ridge National Laboratory, Oak Ridge TN 37831, USA}
\author{T.~T. King}
\affiliation{Physics Division, Oak Ridge National Laboratory, Oak Ridge TN 37831, USA}
\author{B.~C. Rasco}
\affiliation{Physics Division, Oak Ridge National Laboratory, Oak Ridge TN 37831, USA}
\author{K.~P. Rykaczewski}
\affiliation{Physics Division, Oak Ridge National Laboratory, Oak Ridge TN 37831, USA}
\author{H. Schatz}
\affiliation{Department of Physics and Astronomy, Michigan State University, East Lansing, MI 48824, USA}
\affiliation{Facility for Rare Isotope Beams, Michigan State University, East Lansing, MI 48824, USA}
\author{B.~M. Sherrill}
\affiliation{Department of Physics and Astronomy, Michigan State University, East Lansing, MI 48824, USA}
\affiliation{Facility for Rare Isotope Beams, Michigan State University, East Lansing, MI 48824, USA}
\author{O.~B. Tarasov}
\affiliation{Facility for Rare Isotope Beams, Michigan State University, East Lansing, MI 48824, USA}
\author{B. A. Brown}
\affiliation{Department of Physics and Astronomy, Michigan State University, East Lansing, MI 48824, USA}
\affiliation{Facility for Rare Isotope Beams, Michigan State University, East Lansing, MI 48824, USA}
\author{S. Ajayi}
\affiliation{Department of Physics, Florida State University, Tallahassee FL 32306, USA}
\author{H. Arora}
\affiliation{Department of Physics, Central Michigan University, Mt Pleasant MI 48858, USA}
\author{A.~D. Ayangeakaa}
\affiliation{Department of Physics and Astronomy, University of North Carolina, Chapel Hill NC 27599, USA}
\affiliation{Triangle Universities Nuclear Laboratory, Duke University, Durham, North Carolina 27708, USA}
\author{H.~C. Berg}
\affiliation{Department of Physics and Astronomy, Michigan State University, East Lansing, MI 48824, USA}
\affiliation{Facility for Rare Isotope Beams, Michigan State University, East Lansing, MI 48824, USA}
\author{J.~M. Berkman}
\affiliation{Department of Chemistry, Michigan State University, East Lansing, MI 48824, USA}
\affiliation{Facility for Rare Isotope Beams, Michigan State University, East Lansing, MI 48824, USA}
\author{D.~L. Bleuel}
\affiliation{Lawrence Livermore National Laboratory, Livermore CA 94550, USA}
\author{K. Bosmpotinis}
\affiliation{Department of Physics and Astronomy, Michigan State University, East Lansing, MI 48824, USA}
\affiliation{Facility for Rare Isotope Beams, Michigan State University, East Lansing, MI 48824, USA}
\author{M.~P. Carpenter}
\affiliation{Physics Division, Argonne National Laboratory, Argonne IL 60439, USA}
\author{G. Cerizza}
\affiliation{Facility for Rare Isotope Beams, Michigan State University, East Lansing, MI 48824, USA}
\author{A. Chester}
\affiliation{Facility for Rare Isotope Beams, Michigan State University, East Lansing, MI 48824, USA}
\author{J.~M. Christie}
\affiliation{Department of Physics and Astronomy, University of Tennessee, Knoxville, TN 37996, USA}
\author{H.~L. Crawford}
\affiliation{Nuclear Science Division, Lawrence Berkeley National Laboratory, Berkeley CA 94720, USA}
\author{B.~P. Crider}
\affiliation{Department of Physics and Astronomy, Mississippi State University, Mississippi State MS 39762, USA}
\author{J. Davis}
\affiliation{Pacific Northwest National Laboratory, Richland WA 99354, USA}
\author{A.~A. Doetsch}
\affiliation{Department of Physics and Astronomy, Michigan State University, East Lansing, MI 48824, USA}
\affiliation{Facility for Rare Isotope Beams, Michigan State University, East Lansing, MI 48824, USA}
\author{J.~G. Duarte}
\affiliation{Lawrence Livermore National Laboratory, Livermore CA 94550, USA}
\author{A. Estrade}
\affiliation{Department of Physics, Central Michigan University, Mt Pleasant MI 48858, USA}
\author{A. Fijalkowska}
\affiliation{Faculty of Physics, University of Warsaw, Warsaw PL 02-093, Poland}
\author{C. Frantzis}
\affiliation{Department of Physics and Astronomy, Michigan State University, East Lansing, MI 48824, USA}
\affiliation{Facility for Rare Isotope Beams, Michigan State University, East Lansing, MI 48824, USA}
\author{K. Fukushima}
\affiliation{Facility for Rare Isotope Beams, Michigan State University, East Lansing, MI 48824, USA}
\author{T. Gaballah}
\affiliation{Department of Physics and Astronomy, Mississippi State University, Mississippi State MS 39762, USA}
\author{T. Gray}
\affiliation{Physics Division, Oak Ridge National Laboratory, Oak Ridge TN 37831, USA}
\author{E. Good}
\affiliation{Facility for Rare Isotope Beams, Michigan State University, East Lansing, MI 48824, USA}
\affiliation{Pacific Northwest National Laboratory, Richland WA 99354, USA}
\author{K. Haak}
\affiliation{Department of Physics and Astronomy, Michigan State University, East Lansing, MI 48824, USA}
\affiliation{Facility for Rare Isotope Beams, Michigan State University, East Lansing, MI 48824, USA}
\author{S. Hanai}
\affiliation{Center for Nuclear Study, University of Tokyo, Wako, Saitama 351-0198, Japan}
\author{A.~C. Hartley}
\affiliation{Department of Physics and Astronomy, Michigan State University, East Lansing, MI 48824, USA}
\affiliation{Facility for Rare Isotope Beams, Michigan State University, East Lansing, MI 48824, USA}
\author{J.~T. Harke}
\affiliation{Lawrence Livermore National Laboratory, Livermore CA 94550, USA}
\author{K. Hermansen}
\affiliation{Department of Physics and Astronomy, Michigan State University, East Lansing, MI 48824, USA}
\affiliation{Facility for Rare Isotope Beams, Michigan State University, East Lansing, MI 48824, USA}
\author{C. Harris}
\affiliation{Department of Physics and Astronomy, Michigan State University, East Lansing, MI 48824, USA}
\affiliation{Facility for Rare Isotope Beams, Michigan State University, East Lansing, MI 48824, USA}
\author{M. Hausmann}
\affiliation{Facility for Rare Isotope Beams, Michigan State University, East Lansing, MI 48824, USA}
\author{D.~E.~M. Hoff}
\affiliation{Lawrence Livermore National Laboratory, Livermore CA 94550, USA}
\author{D. Hoskins}
\affiliation{Department of Physics and Astronomy, University of Tennessee, Knoxville, TN 37996, USA}
\author{J. Huffman}
\affiliation{Department of Physics and Astronomy, Michigan State University, East Lansing, MI 48824, USA}
\affiliation{Facility for Rare Isotope Beams, Michigan State University, East Lansing, MI 48824, USA}
\author{R. Jain}
\affiliation{Lawrence Livermore National Laboratory, Livermore CA 94550, USA}
\affiliation{Department of Physics and Astronomy, Michigan State University, East Lansing, MI 48824, USA}
\affiliation{Facility for Rare Isotope Beams, Michigan State University, East Lansing, MI 48824, USA}
\author{M. Karny}
\affiliation{Faculty of Physics, University of Warsaw, Warsaw PL 02-093, Poland}
\author{N. Kitamura}
\affiliation{Department of Physics and Astronomy, University of Tennessee, Knoxville, TN 37996, USA}
\author{K. Kolos}
\affiliation{Lawrence Livermore National Laboratory, Livermore CA 94550, USA}

\author{E. Kwan}
\affiliation{Facility for Rare Isotope Beams, Michigan State University, East Lansing, MI 48824, USA}

\author{A. Laminack}
\affiliation{Physics Division, Oak Ridge National Laboratory, Oak Ridge TN 37831, USA}
\author{S.~N. Liddick}
\affiliation{Department of Chemistry, Michigan State University, East Lansing, MI 48824, USA}
\affiliation{Facility for Rare Isotope Beams, Michigan State University, East Lansing, MI 48824, USA}
\author{B. Longfellow}
\affiliation{Lawrence Livermore National Laboratory, Livermore CA 94550, USA}
\author{R.~S. Lubna}
\affiliation{Facility for Rare Isotope Beams, Michigan State University, East Lansing, MI 48824, USA}
\author{S. Lyons}
\affiliation{Pacific Northwest National Laboratory, Richland WA 99354, USA}
\author{M. Madurga}
\affiliation{Department of Physics and Astronomy, University of Tennessee, Knoxville, TN 37996, USA}
\author{M. Mogannam}
\affiliation{Department of Chemistry, Michigan State University, East Lansing, MI 48824, USA}
\affiliation{Facility for Rare Isotope Beams, Michigan State University, East Lansing, MI 48824, USA}
\author{S. Neupane}
\affiliation{Department of Physics and Astronomy, University of Tennessee, Knoxville, TN 37996, USA}
\affiliation{Lawrence Livermore National Laboratory, Livermore CA 94550, USA}
\author{A. Nowicki}
\affiliation{Department of Physics and Astronomy, University of Tennessee, Knoxville, TN 37996, USA}
\author{T.~H. Ogunbeku}
\affiliation{Lawrence Livermore National Laboratory, Livermore CA 94550, USA}
\affiliation{Department of Physics and Astronomy, Mississippi State University, Mississippi State MS 39762, USA}
\affiliation{Facility for Rare Isotope Beams, Michigan State University, East Lansing, MI 48824, USA}
\author{G. Owens-Fryar}
\affiliation{Department of Physics and Astronomy, Michigan State University, East Lansing, MI 48824, USA}
\affiliation{Facility for Rare Isotope Beams, Michigan State University, East Lansing, MI 48824, USA}
\author{J.~R. Palomino}
\affiliation{Department of Physics and Astronomy, Mississippi State University, Mississippi State MS 39762, USA}
\author{M. Portillo}
\affiliation{Facility for Rare Isotope Beams, Michigan State University, East Lansing, MI 48824, USA}
\author{M.~M. Rajabali}
\affiliation{Physics Department, Tennessee Technological University, Cookeville TN 38505, USA}
\author{A.~L. Richard}
\affiliation{Lawrence Livermore National Laboratory, Livermore CA 94550, USA}
\affiliation{Department of Physics and Astronomy, Ohio University, Athens OH 45701, USA}
\author{I. Richardson}
\affiliation{Department of Physics and Astronomy, Michigan State University, East Lansing, MI 48824, USA}
\affiliation{Facility for Rare Isotope Beams, Michigan State University, East Lansing, MI 48824, USA}
\author{E. Ronning}
\affiliation{Department of Chemistry, Michigan State University, East Lansing, MI 48824, USA}
\affiliation{Facility for Rare Isotope Beams, Michigan State University, East Lansing, MI 48824, USA}
\author{G.~E. Rose}
\affiliation{University of California Berkeley, Berkeley CA 94704, USA}
\author{T. Ruland}
\affiliation{Physics Division, Oak Ridge National Laboratory, Oak Ridge TN 37831, USA}
\author{N.~D. Scielzo}
\affiliation{Lawrence Livermore National Laboratory, Livermore CA 94550, USA}
\author{D.~P. Scriven}
\affiliation{Facility for Rare Isotope Beams, Michigan State University, East Lansing, MI 48824, USA}
\author{D. Seweryniak}
\affiliation{Physics Division, Argonne National Laboratory, Argonne IL 60439, USA}
\author{K. Siegl}
\affiliation{Department of Physics and Astronomy, University of Tennessee, Knoxville, TN 37996, USA}
\author{M. Singh}
\affiliation{Department of Physics and Astronomy, University of Tennessee, Knoxville, TN 37996, USA}
\author{M.~K. Smith}
\affiliation{Facility for Rare Isotope Beams, Michigan State University, East Lansing, MI 48824, USA}
\author{A. Spyrou}
\affiliation{Department of Physics and Astronomy, Michigan State University, East Lansing, MI 48824, USA}
\affiliation{Facility for Rare Isotope Beams, Michigan State University, East Lansing, MI 48824, USA}
\author{M. Stepaniuk}
\affiliation{Faculty of Physics, University of Warsaw, Warsaw PL 02-093, Poland}
\author{A. Sweet}
\affiliation{Lawrence Livermore National Laboratory, Livermore CA 94550, USA}
\author{V. Tripathi}
\affiliation{Department of Physics, Florida State University, Tallahassee FL 32306, USA}
\author{A. Tsantiri}
\thanks{Present affiliation: University of Regina}
\affiliation{Department of Physics and Astronomy, Michigan State University, East Lansing, MI 48824, USA}
\affiliation{Facility for Rare Isotope Beams, Michigan State University, East Lansing, MI 48824, USA}
\author{S. Uthayakumaar}
\affiliation{Facility for Rare Isotope Beams, Michigan State University, East Lansing, MI 48824, USA}
\author{W.~B. Walters}
\affiliation{Department of Chemistry, University of Maryland, MD, USA}
\author{S. Watters}
\affiliation{Department of Physics and Astronomy, Michigan State University, East Lansing, MI 48824, USA}
\affiliation{Facility for Rare Isotope Beams, Michigan State University, East Lansing, MI 48824, USA}
\author{M. Wolinska-Cichocka}
\affiliation{Heavy Ion Laboratory, University of Warsaw, PL-02-093 Warsaw, Poland}
\author{R. Yokoyama}
\affiliation{Center for Nuclear Study, University of Tokyo, Wako, Saitama 351-0198, Japan}

\date{\today}

\begin{abstract}
In an experiment performed at the Facility for Rare Isotope Beams (FRIB) 
using the FRIB Decay Station initiator (FDSi), 15 new half lives of 
isotopes near $^{54}$Ca were measured. A new method of extracting lifetimes from experimental data, taking into account the unknown $\beta$-delayed neutron emission branches of very neutron-rich nuclei, was developed to enable systematic uncertainty analysis. The experiment observed a dramatic change in the half-life systematics for the isotopes with neutron number $N$=34. Beyond $N$=34, the decline of nuclear lifetime is much slower, leading to longer than anticipated lifetimes for near-dripline nuclei. State-of-the-art shell-model calculations can explain the experimental results for $Z$$>$19 nuclei, revealing the imprint of shell effects and the need for modification of single-particle neutron states. The results from a newly developed QRPA model with potential for making global predictions were also tested against the experimental results and good agreement was found.
\end{abstract}


\maketitle

\noindent\textbf{\textit{Introduction.}---}
Nuclear $\beta$ decay is a fundamental process for atomic nuclei with highly unbalanced proton-to-neutron ratios, which are the
main focus of many recent experimental and theoretical studies. In discovery experiments on the most unstable species, due to the
lower limit of the required statistics than other spectroscopic information, $\beta$-decay half life ($T_{1/2}$) is often the first
quantity experimentally accessible after the identification of a new isotope \cite{Nakamura2017}. Yet, it provides the crucial
insight into the underlying many-body physics driving the properties of a particular nucleus: the available decay energy, $Q_{\beta}$,
and underlying shell structure effects determine the half lives for neutron-rich nuclei \cite{Kratz_Pfeiffer_Thielemann_1998}. Many
systematic measurements across large groups of isotopes have been carried out at various radioactive ion-beam facilities, aiming to
understand not only nuclear structure but also nucleosynthesis far from the stability \cite{Hosmer2010, Nishimura2021, Quinn2012,
Morales2014, Xu2014, Lorusso2015, Wu2017, Wu2020, Crawford2022}. These measurements are especially valuable close to the drip lines,
where new phenomena are expected to emerge due to weak binding or large proton-neutron asymmetry, and where access to other observables
is very limited due to small production rates and short lifetimes.

The neutron-rich nuclei around $^{54}$Ca attracted a lot of attention in the last decade due to the emergence of the neutron
$N=32,\ 34$ shell closures \cite{Hagen2012, Wienholtz2013, Steppenbeck, Steppenbeck2015, Rosenbusch2015, Garcia2016, Chen2019,
Liu2019, Koszorus2021, Browne2021, Enciu2022, Iimura2023}. The presence of those shell closures in exotic isotopes may impact the
thermal evolution of quiescent $X$-ray binaries as they participate in the Urca cooling, or the electron-capture-$\beta$-decay cycling,
process in the accreted crust \cite{Schatz_2014_Nature}. The exact impact of crust Urca cooling is poorly understood due to the
exoticity of the nuclei involved. Previous studies of Urca cooling in quiescent neutron stars relied on global model predictions,
such as the Quasi-Random Phase Approximation (QRPA) \cite{Schatz_2014_Nature, Deibel_Meisel_Schatz_Brown_Cumming_2016}. Their
predictive powers, however, need to be verified with the experimental $\beta$-decay properties in this region.

In this work, we report new half lives for the neutron-rich nuclei with $Z=17\sim22$ and $N=31\sim41$ measured at the Facility for
Rare Isotope Beams (FRIB) using the FRIB Decay Station initiator (FDSi). A new analysis method is employed to extract lifetimes with
limited statistics. We observe a dramatic change in the systematic behavior of lifetimes that can be explained by state-of-the-art calculations that include shell effects and tensor interactions \cite{Otsuka_2020}. While the current generation of global models fails to reproduce the change in systematics, a next-generation QRPA model was tested against the experimental data and is able to better describe the results.

\begin{figure}
    \centering
    \includegraphics[trim = 5cm 3.5cm 3cm 4cm, clip=true, width=0.45\textwidth]{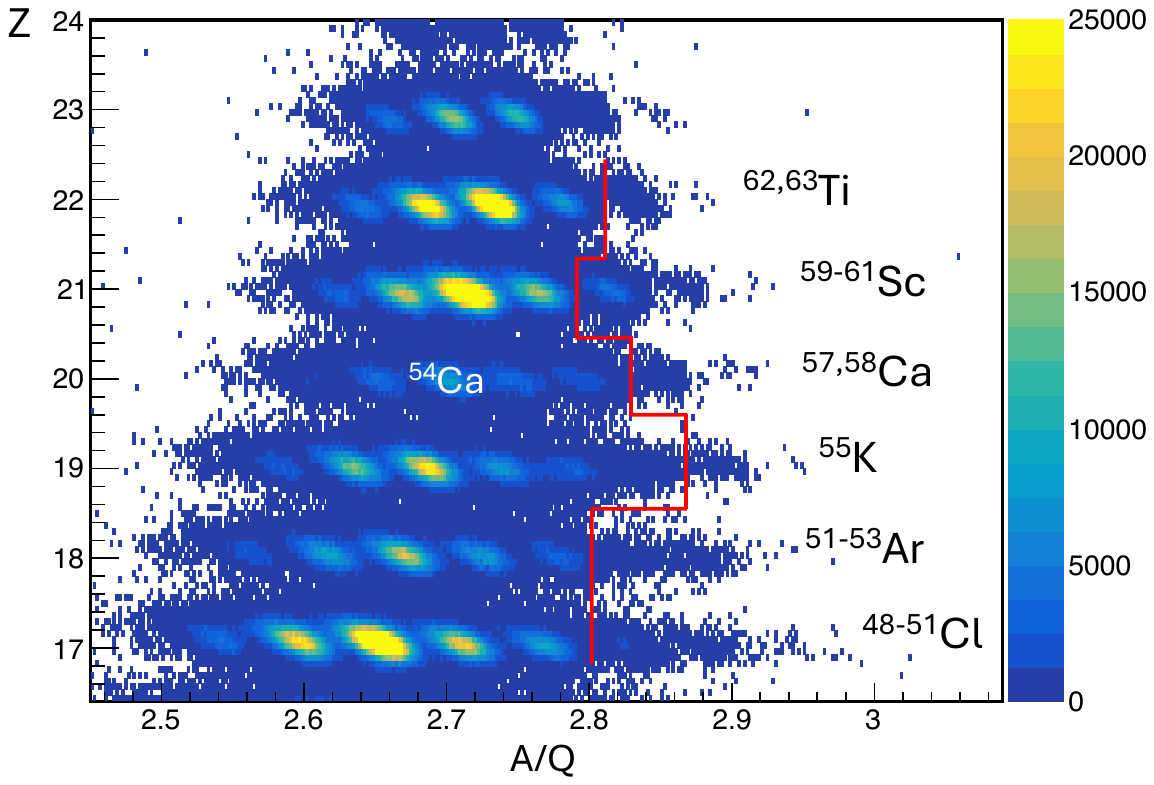}
    \caption{Particle identification plot showing proton number $Z$ and mass-to-charge ratio $A/Q$ for the cocktail beam delivered to the first focal plane of the FDSi. Nuclei with previously unreported half lives are located to the right of the red line. The $A/Q$ = 2.7 nucleus $^{54}$Ca is labeled.}
    \label{fig:PID}
\end{figure}

\noindent\textbf{\textit{Experiment.}---}
The experiment was performed at FRIB in January 2024. The FRIB linear accelerator bombarded a 5-mm C target with a 215~MeV/nucleon, 10-kW $^{82}$Se primary beam. A cocktail beam containing the neutron-rich fragments around $^{54}$Ca was selected by the Advanced Rare Isotope Separator (ARIS) \cite{HAUSMANN2013349, PORTILLO2023151}. Two beam settings were used at ARIS: one centered on $^{55}$Ca (with the magnetic rigidity B$\rho$ = 5.2642~Tm) and the other centered on $^{54}$K (B$\rho$ = 5.3613~Tm). Particle identification (PID) was done on an event-by-event basis using the Energy Loss-Time of Flight ($\Delta E$-TOF) method, where $\Delta E$ was taken from a MSX100 Si PIN detector located upstream of the decay station, and TOF was taken between the anode signal of a parallel-plate avalanche counter at diagnostics box 3 of ARIS (DB3) and a fast plastic scintillator located just downstream of the MSX100 PIN detector. The resultant PID, shown in Fig.\ \ref{fig:PID}, allows for the separation of different isotopes via $Z$ and $A/Q$. 

The radioactive isotopes were stopped at the FDSi, a modular two-focal-plane setup. At the center of the first focal plane, a segmented Yttrium Orthosilicate (YSO) implantation detector \cite{YOKOYAMA201993, SINGH2025170239} is surrounded by 11 HPGe clover detectors, 15 LaBr$_3$ detectors, and the VANDLE array \cite{PETERS2016122}. The second focal plane, downstream of the first focal plane, consists of another YSO-based implantation detector located at the target position of the Modular Total Absorption Spectrometer (MTAS) \cite{KARNY201683}. The YSO detector at the first (second) focal plane was backed by a position-sensitive photomultiplier tube (multi-anode silicon photomultiplier tube) to provide the position information, orthogonal to the beam axis, of the implanted beam particles as well as their subsequent decay electrons. All results from this paper are from the B$\rho$ setting centered on $^{54}$K and taken at the first focal plane.

\noindent\textbf{\textit{Analysis and results.}---}
While spectroscopic data taken by the HPGe and VANDLE detectors will be reported elsewhere, this Letter focuses on fifteen new half lives obtained for $^{48-51}$Cl, $^{51-53}$Ar, $^{55}$K, $^{57,58}$Ca, $^{59-61}$Sc, and $^{62,63}$Ti through decay curve fits. The results are summarized in \cref{table_result}, together with a plot showing their experimental decay curves in \cref{fig:newhalflives} of End Matter. In addition, 12 nuclei with previously reported half lives are also listed in the table, along with the literature values.


To obtain the half lives, each decay curve was fit with the Bateman equations \cite{bateman1910solution}, including the parent, daughter, and granddaughter generations, as well as a flat background. In this region of the nuclear chart, the $Q_{\beta}$ values exceed 10~MeV and the $\beta$-delayed one-neutron ($\beta n$), two-neutron ($\beta2n$), and even three-neutron ($\beta3n$) decays are energetically possible \cite{ame2020}. All fifteen of the nuclei where new half lives are obtained do not have experimentally measured $\beta xn$ branching ratios ($P_{xn}$), and therefore the daughter and granddaughter generations included in the fit also include all the corresponding neutron daughters and neutron granddaughters. In the case where the daughter half life is previously unreported, the fit value from this work was used as input--- for example, the half life obtained for $^{51}$Ar was used in the fit of $^{51}$Cl. To determine the impact of the unknown $P_{xn}$, as well as the uncertainties in the half lives for the daughter and granddaughter decays, a Markov Chain Monte Carlo (MCMC) \cite{neal1993markov} analysis was used. The MCMC method was implemented using the emcee package \cite{foreman-mackey2013emcee} in Python, which utilizes the Metropolis-Hastings algorithm \cite{mackay2003information} to determine the posterior distributions of the fit parameters (e.g., various $T_{1/2}$ or $P_{xn}$). 

\begin{table}[h!]
\renewcommand{\arraystretch}{1.5}
\begin{threeparttable}
 \caption{half lives (in ms) and statistics for nuclei compared to experimental literature values where available, as well as calculated values using the sdpf-mu and ufp-ca interactions. The total number of decay events to which the half-life fit was made is given in the column ``$N_{\beta}$'', and is already corrected for random correlations and the decays of the respective daughter and granddaughter nuclei.}\label{table_result}
\begin{tabularx}{0.5\textwidth}{@{\extracolsep{\fill}}c c c c c c}
  \hline \hline
  
            \multirow{2}{*}{Nuclide} & \multirow{2}{*}{$N_{\beta}$} & \multicolumn{4}{c}{$T_{1/2}$ (ms)}  \\
                     \cline{3-6} 
            &  & This Work & Literature & sdpf-mu & ufp-ca \\
 	 \hline \hline
    $^{47}$Cl	&	7376	& 108.9 $^{+4.3} _{-4.1}$	&	101(6) \cite{GREVY2004252}&	106.5 & $-$ \\
    $^{48}$Cl   &	2401	&	 36.6 $^{+2.9} _{-2.0}$	&		& 42.6   & $-$ \\ 
    $^{49}$Cl	&	442	&	35.7 $^{+6.2} _{-5.0}$	&		& 36.2  & $-$ \\
    $^{50}$Cl	&	73	&	17.3 $^{+5.0} _{-3.9}$	&		&  20.8 & $-$ \\
    $^{51}$Cl	&	16	&	8.2 $^{+4.4} _{-2.7}$	&		&  9.0 &$-$  \\
     $^{50}$Ar	&	3185	& 128.6 $^{+13.0} _{-12.6}$	&	106(6) \cite{Weissman_2012}& 127.5 & $-$ \\
    $^{51}$Ar	&	232	&	48.9 $^{+12.1} _{-9.9}$	&		&	45.7 & $-$ \\
    $^{52}$Ar	&	552	&	22.6 $^{+3.5} _{-3.1}$	&		& 19.6 & $-$ \\
    $^{53}$Ar	&	16	&	6.5 $^{+8.0} _{-5.4}$	&		& 5.6 & $-$ \\
     $^{53}$K	&	2021	& 43.3 $^{+2.6} _{-2.6}$	&	30(5) \cite{Langevin_PLB_1983}&	38.4 & $-$ \\
     $^{54}$K	&	2319	& 13.3 $^{+0.6} _{-0.6}$	&	10(5) \cite{Langevin_PLB_1983}&	8.5 & $-$ \\
    $^{55}$K	&	498	&	6.4 $^{+0.7} _{-0.6}$	&		&  3.4 & $-$ \\
    $^{55}$Ca	&	15418	& 21.5 $^{+1.0} _{-0.8}$	& 22(2)\cite{Mantica2008} & 14.7 & 18.1 \\
     $^{56}$Ca	&	9838	& 12.9 $^{+0.7} _{-0.6}$	&	11(2) \cite{MANTICA_Ca56}& 6.7 & 7.8 \\
    $^{57}$Ca	&	854	&	6.1 $^{+0.7} _{-0.5}$	&		& 4.3 & 4.4 \\
    $^{58}$Ca	&	36	&	5.2 $^{+2.9} _{-2.5}$	&		& 3.4 & 3.2 \\
    $^{57}$Sc $^a$	&	142	&	18.3$^{+0.3} _{-0.3}$	& 13(4) \cite{Gaudefroy2004}	& 11.8 & 15.7 \\
     $^{58}$Sc	&	60052	& 10.1 $^{+0.1} _{-0.1}$	&	12(5) \cite{Gaudefroy2005}&	9.1 & 10.5 \\
    $^{59}$Sc	&	11079	&	7.5$^{+0.3} _{-0.2}$	&		& 4.6 & 5.0 \\
    $^{60}$Sc	&	278	&	6.2 $^{+0.8} _{-0.7}$	&		& 3.2 & 3.8 \\
    $^{61}$Sc	&	17	&	2.8 $^{+2.6} _{-2.0}$	&		& 5.1  & 2.4 \\
     $^{58}$Ti	& 10593  & 57.7 $^{+4.9} _{-3.8}$	&	58(9) \cite{Daugas2011} & 47.5 & 35.8 \\
    $^{59}$Ti	&	66424	& 32.8 $^{+1.5} _{-1.2}$	&	28(3) \cite{Daugas2011}, 30(3) \cite{Gaudefroy2005} & 23.7 & 23.0 \\
   
    $^{60}$Ti $^b$ & 145	& 22.3 $^{+0.6} _{-0.6}$	&	22(3) \cite{Daugas2011}, 22(2) \cite{Gaudefroy2005} & 12.5 & 12.1 \\
     $^{61}$Ti	&	14566 & 15.1 $^{+0.3} _{-0.2}$	&	15(4) \cite{Daugas2011}& 7.1 & 7.3 \\
    $^{62}$Ti	&	1510	&	9.9 $^{+0.6} _{-0.6}$	&	 & 5.5 & 4.6 \\
    $^{63}$Ti	&	14	&	3.6 $^{+5.1} _{-2.6}$	&		&	$-$ & $-$ \\
   \hline
\end{tabularx}
\begin{tablenotes}
\item[a] Gated on 364 keV $\gamma$ ray \cite{Crawford_2010}
\item[b] Gated on previously unreported 111 keV $\gamma$ ray
\end{tablenotes}
\end{threeparttable}
\end{table}

Flat prior distributions were assumed for the parent half life and implanted activity. The flat background component was constrained by using a Gaussian distribution with the mean and standard deviation from the $T < 0 $ portion of the decay curves. The priors for the $\beta$-decay half lives of the daughter and granddaughter generations were Gaussian distributions with means and standard deviations of the most recently published or evaluated data, while the individual $P_{xn}$ priors were assumed to be uniform between 0 and 1, with the sum of all branches enforced to be unity ($\Sigma_{j=0}^{j=n}P_{jn} = 1$). For low statistics cases, the log-likelihood method was employed, while for higher statistics cases, where the two methods converge, the $\chi^2$ method was used. The first 25$\%$ of iterations were discarded to ensure convergence of the MCMC.

\begin{figure}[htbp]
    \centering
    \includegraphics[width=0.45\textwidth]{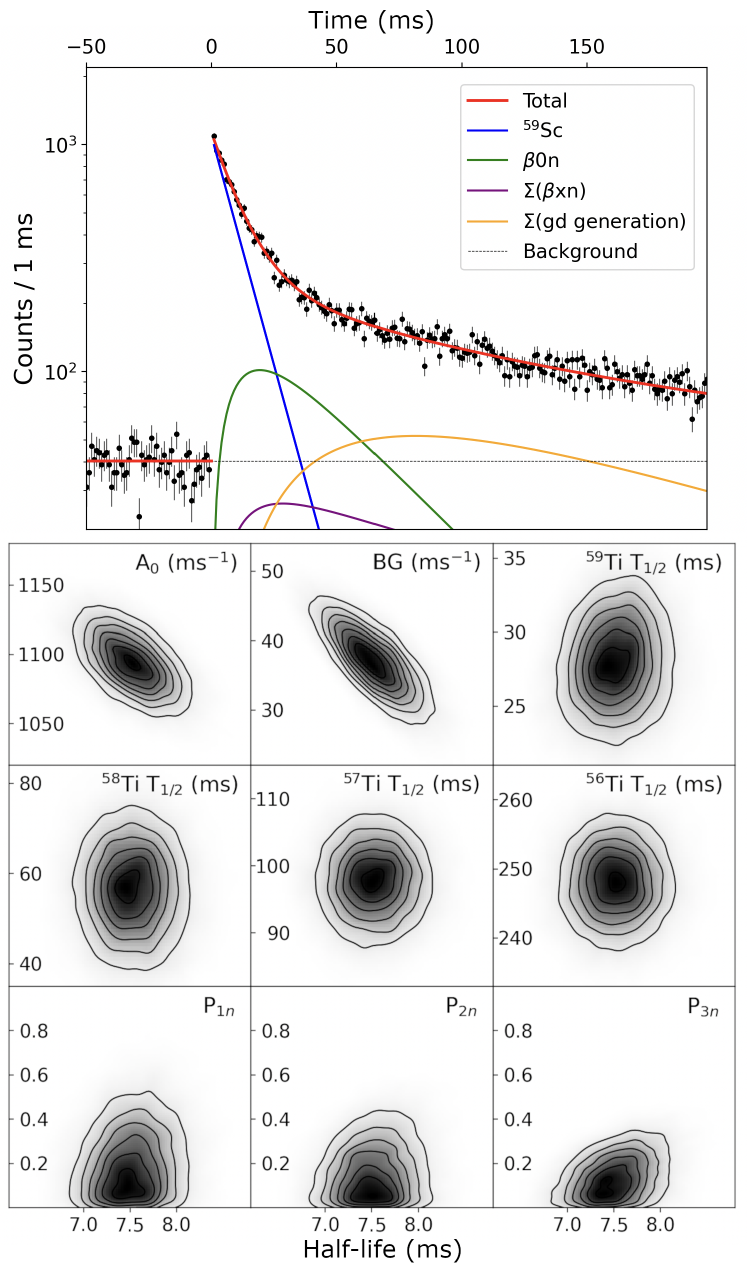}
    \caption{Top: Decay curve $^{59}$Sc showing the different components of the fit. Bottom: The 2-D posterior distributions of the fit parameters for $^{59}$Sc.}
    \label{fig:newfig2_Oct2}
\end{figure}
As an example, the analysis results for $^{59}$Sc, including the fitted decay curve and the posterior distributions, are shown in Fig. \ref{fig:newfig2_Oct2}.
The MCMC results show that the fit result for the parent half life is most correlated to the activity at the peak of the decay curve ($A_0$). Though there is some correlation with the background value, the strict constraint imposed by the negative time portion of the decay curve means that the contribution of the background to the total uncertainty is minimal. Importantly, the correlations between the parent half life and the unknown $P_{xn}$ branches are relatively weak, with typical Pearson's correlation coefficients of $-0.25 < \rho < 0.25$. Lastly, the daughter and granddaughter half lives show some marginal correlation with the fit parent half life.
We also considered the error associated with the choice of bin width, and the beginning and ending points of the fit. This was done with Monte Carlo, where the MCMC fitting procedure itself was repeated while varying the three parameters through a grid of values.



\begin{figure*}[htpb]
   \centering
   \includegraphics[width=1.00\textwidth]{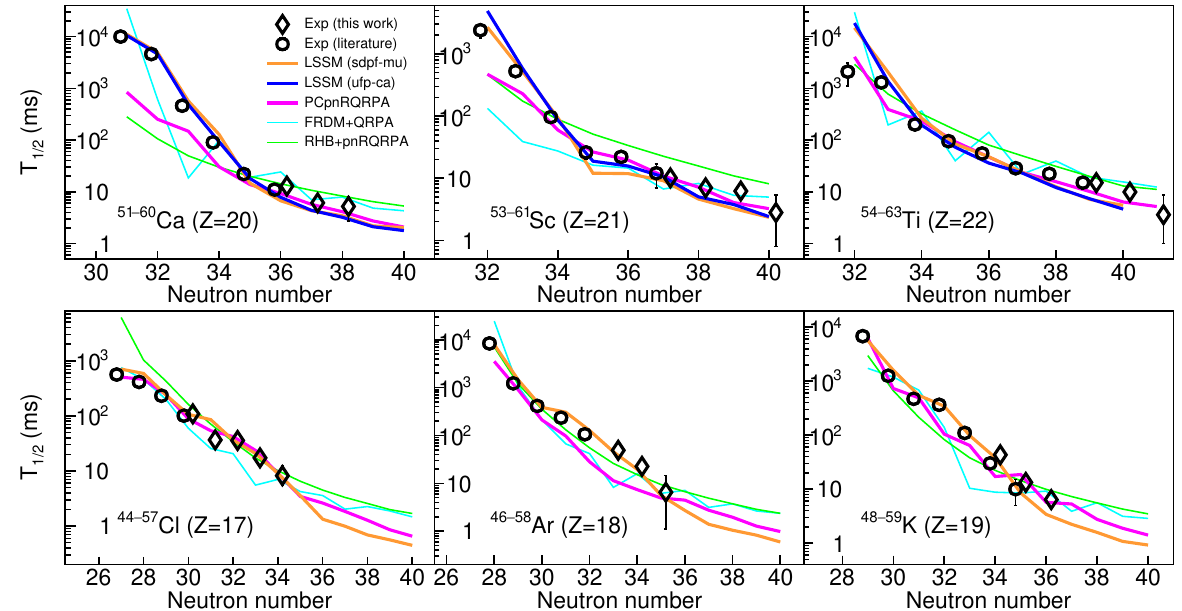}
   \caption{
      Half-life systematics of neutron-rich isotopes from chlorine ($Z=17$) to
      titanium ($Z=22$). The experimental data are compared with FRDM+QRPA
      \cite{moller2019}, RHB+pnRQRPA \cite{marketin2016}, PCpnRQRPA, and LSSM
      with the sdpf-mu and ufp-ca interactions.
   }\label{e21069b_sys}
\end{figure*}

\noindent\textbf{\textit{Discussion.}---}
To compare with our measurements and provide insights into the underlying
shell structure, we computed $\beta$-decay half lives using large-scale
shell models (LSSM). Firstly, we employed the sdpf-mu interaction \cite{sdpf-mu},
which has been proven to be very successful in this region (see, e.g., \cite{cox24} and
references therein). Its $sd$-$pf$ model space covers all the isotopes of
interest with $Z=17\sim22$ in this work except for $^{63}$Ti. Similar to the
previous studies \cite{yoshida18, cox24}, the number of particle-hole excitations
between the $sd$ and $pf$ shells is limited to one ($\le1\hbar\omega$) in our calculations. From the initial ground state, both the allowed
Gamow-Teller (GT) and first-forbidden (FF) transitions were considered, with the
same effective scaling factors as Ref. \cite{yoshida18} for the respective transition
operators. In addition, we performed half-life calculations using the newly developed ufp-ca Hamiltonian \cite{Magilligan2021}. Because its model space is limited to the $pf$ shell, the calculations were performed for isotopes with $Z=20\sim22$ without extra truncation, and only the GT channels were included, with a generic 0.75 quenching factor of the GT operator \cite{caurier2005}. The ground-state spins and parities ($I^{\pi}$) of parent nuclei were taken from experimental data whenever available. Otherwise, we used the values from both LSSM predictions. The strength distributions were converted to half lives using the atomic masses available in AME2020 \cite{ame2020}. For more exotic isotopes
beyond the AME2020 table, we used the $Q_{\beta}$ values from our LSSM calculations. The results are listed in Table \ref{table_result}.

\begin{figure}[htbp]
    \centering
    \includegraphics[width=0.5\textwidth,trim=9.6cm 1cm 10cm 2cm,clip=true]{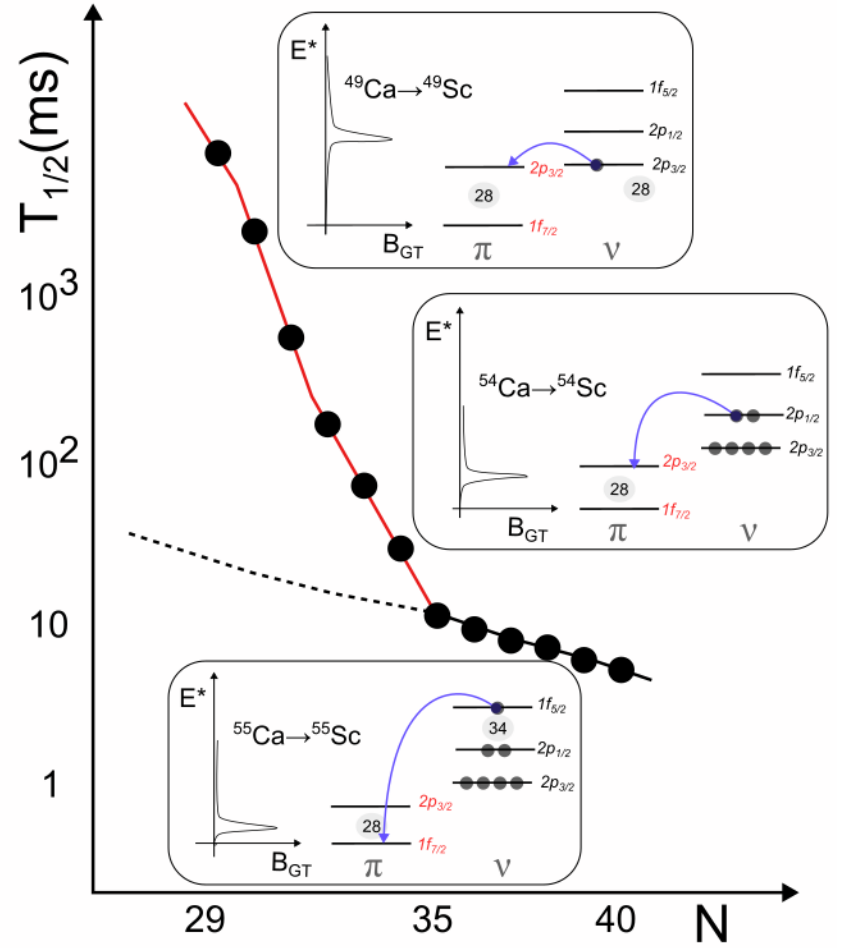}
    \caption{
    Schematic drawing of the dominant GT channels in the systematics of $^{49-60}$Ca.} 
    \label{configsp}
\end{figure}

\Cref{e21069b_sys} presents the comparison between the calculated results and the experimental data. We also plot predictions using the
global models FRDM+QRPA \cite{moller2019} and RHB+pnRQRPA \cite{marketin2016}. Our LSSM calculations, with both the sdpf-mu and
ufp-ca interactions, have an overall excellent agreement with the experimental half lives. Most interestingly, they reproduce the striking
kink that emerges in the data. For the calcium ($Z=20$) and scandium ($Z=21$) isotopic chains, the decrease in half lives is significantly
slowed down beyond $N=35$ with increasing neutron number. The diagrams in Fig. 5 illustrate how the shell effects can explain the observed
half-life systematics. According to the LSSM predictions, when the neutron number is increased from $N=28$ towards $N=34$, decays are
dominated by either $\nu p_{1/2}\rightarrow\pi p_{3/2}$ or $\nu p_{3/2}\rightarrow\pi p_{3/2}$. Both transitions populate the states with a
proton in the $p_{3/2}$ orbital above $f_{7/2}$, with the latter at the Fermi surface. Due to the attractive proton-neutron interactions,
those states gradually become lower in excitation energy when more neutrons are added to the $\nu p_{3/2}$ and $\nu p_{1/2}$ orbitals
\cite{otsuka2010}. This, together with the increasing $Q_{\beta}$, amplifies the $\beta$-decay phase space and decreases the half lives
rapidly. Beyond $N=34$, two mechanisms can affect the half-life systematics. First, neutrons start to occupy the $\nu f_{5/2}$ orbital, and
the dominant GT decay channel shifts to $\nu f_{5/2}\rightarrow\pi f_{7/2}$. Since both the proton and neutron orbitals are at the Fermi
surface, the strongest populated states remain at low excitation energy. At the same time, the change from a particle-particle to a
particle-hole configuration above $N=34$ raises the excitation energy of the populated states (e.g., from near zero in $^{54,56}$Sc to $\sim1$
MeV in $^{60}$Sc). These both hamper the growth of the $\beta$-decay phase space, and the impact of increasing B(GT) and $Q_{\beta}$ is thus
lessened. The new half lives confirm the trends in the calcium and scandium isotopes until $N=38$ and $N=40$, respectively. They underline the
significance of properly modeling the proton-neutron interactions and shell structure in order to explain the experimental half-life systematics.

Below $Z=20$, the experimental half lives cannot yet confirm the kink, which is predicted at $N=36$ along $Z=17,19$ and at $N=37$ along $Z=18$.
Here, an explanation for the kink is more complex due to the fact that the proton Fermi surface is down into the $sd$ shell instead of at
$\pi f_{7/2}$, and the FF transitions from a neutron in the $pf$ shell to a proton in the $sd$ shell (with parity change) become strong enough
to compete with the GT transitions and alter the half-life systematics. On the other hand, the new half lives of $^{51}$Cl and $^{53}$Ar, which
might be at or very close to the neutron dripline  \cite{Tarasov2018}, smoothly follow the trends established by the less exotic isotopes. This
suggests the nuclear structure effects emerging at the neutron dripline, such as the extended radial wavefunctions and the coupling to the
continuum, do not induce strong and distinguished anomalies in the ground-state half lives in this region.

Although the LSSM calculations agree well with the experimental data, it is impractical to expand across the nuclear chart for modeling
nucleosynthesis processes in astrophysical environments. We used the half lives measured in this work to benchmark a newly developed global model,
PCpnRQRPA, see \cref{e21069b_sys}. Compared with FRDM+QRPA and RHB+pnRQRPA, the new calculations provide much more consistent predictions. Its
framework is based on the relativistic density functional theory developed in Refs.\ \cite{Ravlic2024, Ravlic2025}. Here we utilize the DD-PCX
interaction \cite{Yuksel2019} with time-odd terms as in Ref.\ \cite{Ravlic2024}, with reduced isoscalar pairing strength $V_0^{is}=1.0$, and
quenched axial-vector coupling $g_A = 0.67$. Under time-reversal symmetry, the particle-hole channel of the EDF can be split into time-even and
time-odd parts. While time-even couplings are constrained by ground-state observables such as radii and masses, $\beta$-decay half-lives probe
charge-changing time-odd terms \cite{Schunck2010, Mustonen2016} and the proton-neutron pairing, of which the strength has been shown to vary
along isotopic chains as nuclei become more neutron rich \cite{Niu2013, Bai2014, Minato2022}. The half lives uncovered in this work could
improve constraints on both the strength of isoscalar pairing as well as the time-odd couplings, providing benchmarks for future generations of
EDFs with enhanced extrapolation towards the neutron dripline.

\noindent\textbf{\textit{Conclusions.}---}
In summary, we report new half-life results from the FDSi experiment focusing on the $Z\sim20,\ N=34$ region. Fifteen new half lives were measured
in this work, and we demonstrated that the systematic error from the unknown $P_{xn}$ is minimal. Our results show good agreement with LSSM
calculations using both the sdpf-mu ($17\le Z\le22$) and ufp-ca ($20\le Z\le22$) interactions. Critically, our measured half lives of $^{57,58}$Ca
and $^{59-61}$Sc confirm the existence of the ``kink'' at $N=35$ along the calcium and scandium isotopic chains. Overall, they provide valuable
feedback to constrain nuclear theories, not only LSSM but also global models, for a better description of $\beta$-decay properties in this region of
the nuclear chart.

\begin{acknowledgments}
This material is based upon work supported by the U.S. Department of Energy, Office of Science, Office of Nuclear Physics, and used resources of the Facility for Rare Isotope Beams (FRIB) Operations, which is a DOE Office of Science User Facility under Award Number DE-SC0023633.
This work was performed under the auspices of the U.S. Department of Energy by Lawrence Livermore National Laboratory under Contract DE-AC52-07NA27344 (LLNL). It was also supported by: the U.S. Department of Energy, Office of Science, Office of Nuclear Physics, under contracts No. DE-AC02-06CH11357 (ANL), No. DE-AC05-00OR22725 (ORNL) and grants No. DE-FG02- 96ER40983 (UTK), DE-FG02-94ER40834 (Maryland), and No. DE-SC0016988 (TTU); the US National Science Foundation under grants No. PHY-23-10078, No. PHY-2209429, No. PHY-2110365; the National Nuclear Security Administration under Award No. DE-NA0003180 and the Stewardship Science Academic Alliances program through DOE Awards No. DOE-DE-NA0003906, DOE-DE-NA0004068, and DOE-DE-NA0004074; the National Science Foundation Major Research Instrumentation Program Award No. 1919735 (UTK, TTU); the Laboratory Directed Research and Development Program at Pacific Northwest National Laboratory operated by Battelle for the U.S. Department of Energy; the Laboratory Directed Research and Development Program at Lawrence Livermore National Laboratory (23-LW-25); the Japan Society for the Promotion of Science, Grants-in-Aid for Scientific Research (KAKENHI) under Grant No. 22K14053. \end{acknowledgments}


\bibliography{apstemplate.bib}

\onecolumngrid
\vspace{1em}
\begin{center}
    \textbf{End Matter}
\end{center}

\begin{figure}[htbp]
    \centering
    \includegraphics[width=\textwidth, trim = 2cm 1.5cm 2cm 0, clip=True]{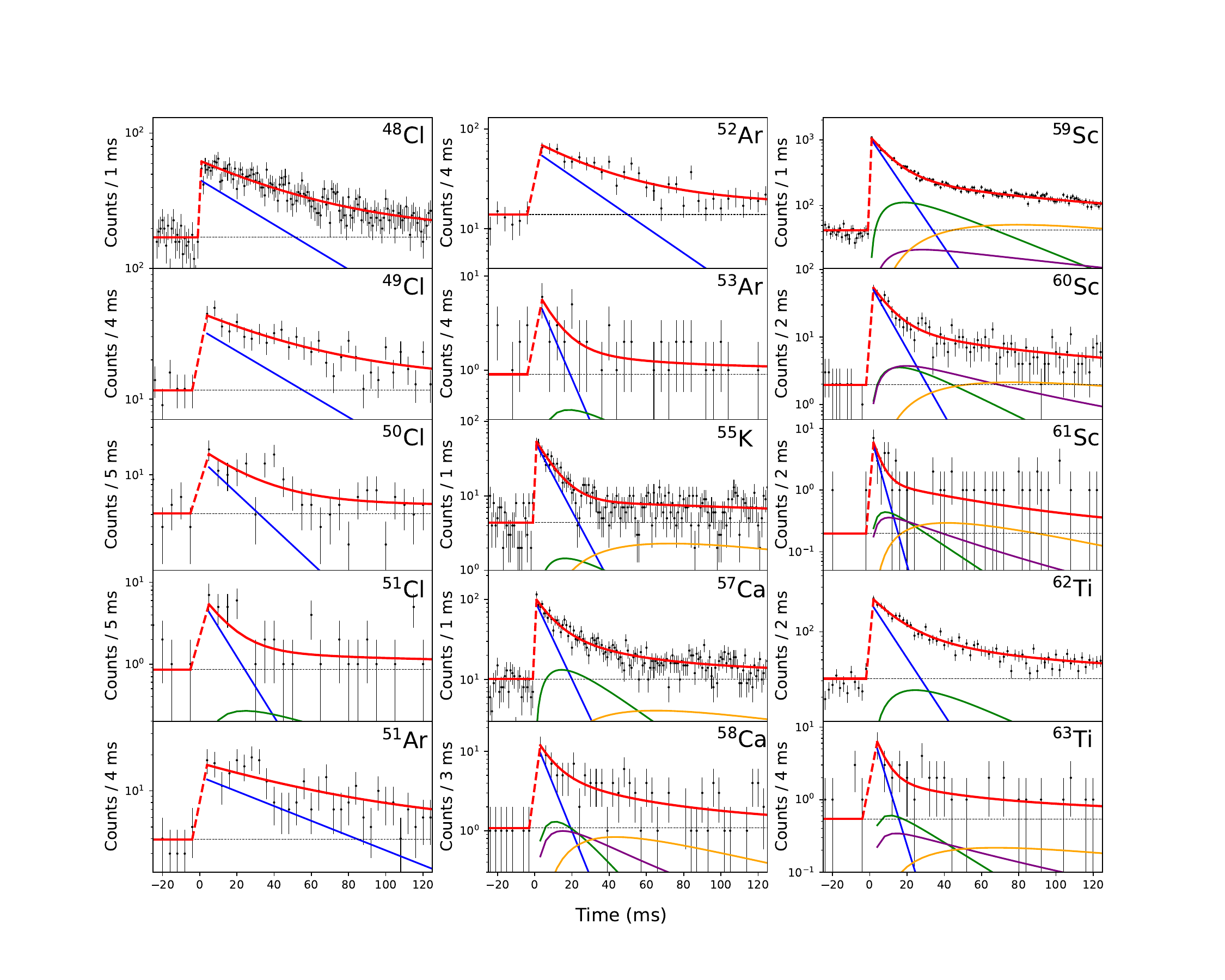}
    \caption{Decay curves for $^{48-51}$Cl, $^{51-53}$Ar, $^{55}$K, $^{57,58}$Ca, $^{59-61}$Sc, and $^{62,63}$Ti. The fits include the decays of the parent nucleus (solid blue), daughter (green solid), sum of neutron daughters (purple solid), and granddaughter (orange solid). The flat background from random correlations is obtained from the time-inverse correlations (black dashed). The total fit is given by the red solid line.}
    \label{fig:newhalflives}
\end{figure}

\end{document}